\documentclass[twocolumn,showpacs,preprintnumbers,amsmath,amssymb]{revtex4}
\usepackage{graphicx}
\usepackage{dcolumn}
\usepackage{bm}

\begin{document}

\title{The medium in heavy-ion collisions}

\author{I.M. Dremin}
 \email{dremin@lpi.ru}
\affiliation{%
Lebedev Physical Institute, \\Moscow 119991, Russia
}%

\date{\today}

\begin{abstract}
The nuclear index of refraction, the density of partons, their free path 
length and energy loss in the matter created in heavy-ion collisions at 
RHIC are estimated within the suggestion that the emission of Cherenkov 
gluons is responsible for the observed two-bump structure of the angular 
distribution of hadrons belonging to the companion (away-side) jet.
\end{abstract}

\pacs{12.38.Mh, 25.75.Dw}

\maketitle

Collective effects and jets are the main tools for studies of the nuclear 
matter created in heavy-ion collisions. The "in-vacuum" and "in-medium" 
jets are different. The "in-vacuum" jets have been well studied in 
$e^+e^-$-collisions (for reviews see, e.g., \cite{koch, dgar}). Jet 
characteristics are, however, modified if jets traverse the nuclear medium. 
Jet quenching or the suppression of high $p_T$ hadron spectra is one of the 
well known effects induced by the dense matter. Elliptic flow and jet quenching 
have been widely used for analysis of partonic properties of the matter, in
particular, for estimations of parton densities \cite{hein, wawa, vgyu, xnwa}.

A quark traversing this medium can emit gluons by the mechanism analogous 
to emission of Cherenkov photons in ordinary media. Thus, Cherenkov gluons 
\cite{d1, d2, dim, mwa, kmwa, stoc, mrup} can serve as another diagnostic 
tool. Similarly to photons, Cherenkov gluons are emitted along the cone. 
Its half-angle is determined by the nuclear index of refraction $n$ as
\begin{equation}
\cos \theta_c=\frac {1}{n}.   \label{cos}
\end{equation}
Therefore, the rings of hadrons similar to usual Cherenkov rings can be
observed in the plane perpendicular to the cone axis if $n>1$.

Recent experimental observations at RHIC \cite{wang} shown in Fig. 1 revealed
the two-bump structure of the angular distribution of hadrons belonging to
the so-called companion (away-side) jet in central heavy-ion collisions. 
There is no such structure in pp-collisions. The difference has been 
attributed to "in-medium" effects. These features are clearly seen in Fig. 1 
(the upper part for pp, lower one for Au-Au). One easily notices the 
remarkable difference between particle distributions in the direction 
opposite to the trigger jet maximum which is positioned at 
$\Delta \phi = 0$. Both trigger and companion high-$p_T$ 
jets have been created in central Au-Au collisions at $\sqrt s=200$ GeV 
at the periphery of a nucleus. They move in opposite directions. The trigger
parton immediately escapes the nucleus and, therefore, is detected as the
"in-vacuum" jet. The companion (away-side) jet traverses the whole nucleus
before it comes out. It is modified by "in-medium" effects. Beside normal
fragmentation, its initiating parton can emit Cherenkov gluons which produce
a ring of hadrons in the plane perpendicular to the jet axis.

Thus we can ascribe two contributions to the away-side hadrons associated
with the companion jet: one from jet fragmentation and the other from Cherenkov
gluons. The hadrons from jet fragmentation are smoothly distributed within the
phase space volume. In distinction, the one-dimensional distribution along the 
ring diameter of the away-side hadrons created by Cherenkov gluons must possess 
two peaks because it is just the projection of the ring on its diameter. The 
distance between these peaks is exactly equal to the diameter.

In angular variables, the ring radius is given by $\theta _c$ in Eq. 
(\ref{cos}). Let us note that $\Delta \phi $ in Fig. 1 coincides with 
$\theta $ in our notations. Herefrom, it is determined that 
$\theta _c \approx 70^0$ or $\langle n\rangle \approx 3$, where the average
should be done over gluon energy if $n$ depends on it.

\begin{figure}
\includegraphics[width=\columnwidth]{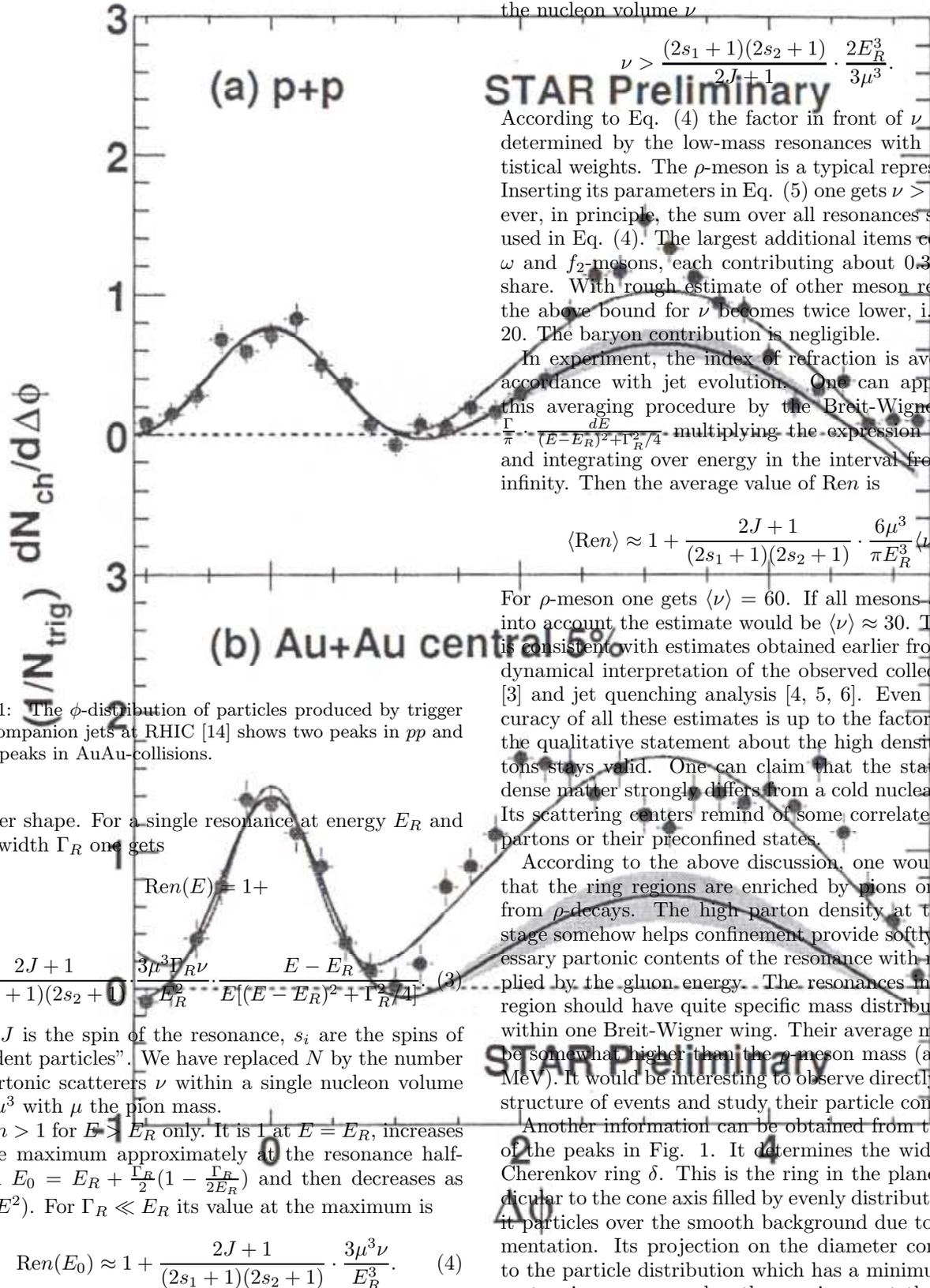}
\caption{
The $\phi $-distribution of particles produced by trigger and 
companion jets at RHIC \cite{wang} shows two peaks in $pp$ and three 
peaks in AuAu-collisions.
}
\end{figure}

For photons, the index of refraction is usually related to their forward
scattering amplitude. We use the similar relation for the nuclear index of
refraction $n$ using the forward scattering amplitude $F(E)$ at energy $E$ 
for gluons:
\begin{equation}
n(E)=1+\frac {2\pi N}{E}F(E).       \label{ne}
\end{equation}
Here $N$ is the density of the scattering centers in a nucleus. The necessary
condition for emission of Cherenkov gluons is ${\rm Re} n(E)>1$ according to Eq.
(\ref{cos}). QCD estimates of $F(E)$ are rather indefinite. Therefore, we 
rely on general properties of hadronic reactions known from experiment. The 
real part of hadronic scattering amplitudes becomes positive either within 
the upper wing of a Breit-Wigner resonance or at very high energies. Jet 
energies available in RHIC experiment \cite{wang} (Fig. 1) are sufficient 
only for production of resonances. Therefore, we attribute
the feature shown in Fig. 1 to resonance effects as discussed in \cite{dim}.
If the hadronization of gluons is a soft process then the gluon energy closely
corresponds to the energy of the produced resonance. It implies that in this
particular experiment \cite{wang} Cherenkov gluons can be emitted only 
with energies within the upper wings of hadronic resonances. Their amplitude 
is of the Breit-Wigner shape. For a single resonance at energy $E_R$ and 
with width $\Gamma _R$ one gets
$$
{\rm Re} n(E)=1+
$$
\begin{equation}
\frac {2J+1}{(2s_1+1)(2s_2+1)}\cdot \frac {3\mu ^3\Gamma _R\nu }
{E_R^2}\cdot \frac{E-E_R}{E[(E-E_R)^2+\Gamma _R^2/4]}.    \label{ren}
\end{equation}
Here $J$ is the spin of the resonance, $s_i$ are the spins of "incident 
particles". We have replaced $N$ by the number of partonic scatterers $\nu $
within a single nucleon volume $4\pi /3\mu ^3$ with $\mu $ the pion mass.

${\rm Re} n>1$ for $E>E_R$ only. It is 1 at $E=E_R$, increases to the maximum 
approximately at the resonance half-width
$E_0=E_R+\frac {\Gamma _R}{2}(1-\frac {\Gamma _R}{2E_R})$
and then decreases as $O(1/E^2)$. For $\Gamma _R\ll
E_R$ its value at the maximum is
\begin{equation}
{\rm Re} n(E_0)\approx 1+\frac {2J+1}{(2s_1+1)(2s_2+1)}\cdot \frac {3\mu ^3\nu }
{E_R^3}.                     \label{ren0}
\end{equation}
The average value of ${\rm Re} n$ is smaller than its value at the maximum. 
Therefore, from experimental estimate of this average value equal to 3 and
Eq. (\ref{ren0}) one gets the lower bound for the effective number of partons 
within the nucleon volume $\nu $
\begin{equation}
\nu >\frac {(2s_1+1)(2s_2+1)}{2J+1}\cdot \frac {2E_R^3}{3\mu ^3}.  \label{num}
\end{equation}
According to Eq. (\ref{ren0}) the factor in front of $\nu $ is mostly 
determined by the low-mass resonances with high statistical weights.
The $\rho $-meson is a typical representative. Inserting its parameters
in Eq. (\ref{num}) one gets $\nu > 40$. However, in principle, the sum over 
all resonances should be used in Eq. (\ref{ren0}). The largest additional
items come from $\omega $ and $f_2$-mesons, each contributing about 0.3 of 
the $\rho $ share. With rough estimate of other meson resonances the above
bound for $\nu $ becomes twice lower, i.e. about 20. The baryon contribution
is negligible.

In experiment, the index of refraction is averaged in accordance with jet
evolution. One can approximate this averaging procedure by the Breit-Wigner
weight $\frac {\Gamma }{\pi }\cdot \frac {dE}{(E-E_R)^2+\Gamma _R^2/4}$ 
multiplying the expression (\ref{ren}) by it and integrating over energy in 
the interval from $E_R$ to infinity. Then the average value of ${\rm Re} n$ is
\begin{equation}
\langle {\rm Re} n\rangle \approx 1+\frac {2J+1}{(2s_1+1)(2s_2+1)}\cdot \frac 
{6\mu ^3 }{\pi E_R^3}\langle \nu \rangle.    \label{nuav}
\end{equation}
For $\rho $-meson one gets $\langle \nu \rangle=60$. If all mesons are taken 
into account the estimate would be $\langle \nu \rangle \approx 30$. This
value is consistent with estimates obtained earlier from hydrodynamical
interpretation of the observed collective flow 
\cite{hein} and jet quenching analysis \cite{wawa, vgyu, xnwa}. Even if 
the accuracy of all these estimates is up to the factor about 2, the 
qualitative statement about the high density of partons stays valid. One can
claim that the state of this dense matter strongly differs from a cold nuclear
matter. Its scattering centers remind of some correlated current partons or
their preconfined states.

According to the above discussion, one would expect that the ring regions 
are enriched by pions originating from $\rho $-decays. The high parton density
at the initial stage somehow helps confinement provide softly the necessary
partonic contents of the resonance with mass supplied by the gluon energy.
The resonances in the ring region should have quite specific mass distribution 
only within one Breit-Wigner wing. Their average mass must be somewhat higher 
than the $\rho $-meson mass (about 840 MeV). It would 
be interesting to observe directly the ring structure of events and study their 
particle contents.

Another information can be obtained from the height of the peaks in Fig. 1.
It determines the width of the Cherenkov ring $\delta $. This is the ring 
in the plane perpendicular to the cone axis filled by evenly distributed 
within it particles over the smooth background due to jet fragmentation.
Its projection on the diameter corresponds to the particle distribution which
has a minimum at the center, increases, reaches the maximum at the internal
radius of the ring $r_1$ and then decreases to zero at its external radius 
$r_2$. For narrow rings ($\delta \ll r_i$) the height of the maximum over the
minimum is easily determined as
\begin{equation}
h_{max}=\sqrt {2r_1\delta }-\delta .     \label{max}
\end{equation}
With $h_{max} \approx 1.6 - 1.2 = 0.4$ and $r_1 = 1.2$ in Fig. 1 one gets
\begin{equation}
\delta \approx 0.1.        \label{delt}
\end{equation}
The ring of Cherenkov gluons is really quite narrow. Actually, Eq. (\ref{cos}) 
implies that the ring is squeezed to a circle.

There are three physical reasons which can lead to the finite width of the ring. 
First, it is the dispersion, i.e. the energy dependence of the index of 
refraction. Its contribution to the width is well known \cite{land}
\begin{equation}
\delta _d=\int _0^{\delta _d}d\theta =\cot \theta _c\int _0^{\infty }
\frac {1}{n} \frac{dn}{dE} dE.      \label{lan}
\end{equation}
If the Breit-Wigner expression (\ref{ren}) for $n(E)$ is used, the result is
\begin{equation}
\delta _d=0.   \label{deld}
\end{equation}
It is amazing that there is no widening of the Cherenkov cone due to the 
dispersion of $n(E)$ described by the formula (\ref{ren}) with Breit-Wigner
resonances.

Second, the width of the Cherenkov ring can be due to the finite free path 
length for partons. Qualitatively, it can be estimated as the ratio of the 
parton wavelength $\lambda $ to the free path length $R_f$
\begin{equation}
\delta _f \sim \frac {\lambda }{R_f}.    \label{rf}
\end{equation}
For $\lambda \sim 2/E_R$ and $\delta _f < 0.1$ one gets the estimate for
the free path length
\begin{equation}
R_f>20/E_R \sim 4.5/\mu \sim 7\cdot 10^{-13} cm.   \label{rfn}
\end{equation}
This appears to be quite a reasonable estimate.

Finally, the width of the ring can become larger due to resonance decays.
That is why there is the inequality sign in (\ref{rfn}). However, this can be 
quantified only if the Monte Carlo program for jets with Cherenkov gluons is
elaborated.

Let us note that the contribution of Cherenkov gluon emission to the away-side
hadrons at the minimum in Fig. 1 is rather small, of the order of 0.1 compared
to the total value at the minimum about 1.2. Jet fragmentation contributes an
order of magnitude larger number of hadrons at this point.

The energy loss can be calculated using the standard formula
\begin{equation}
\frac {dE}{dx}=4\pi \alpha _S\int _{E_R}^{E_R+\Gamma _R}E\left (1-\frac {1}
{n^2(E)}\right )dE\approx 1 GeV/fm.       \label{dedx}
\end{equation}
This estimate is an order of magnitude higher than the value of 0.1 GeV/fm 
obtained in the model of \cite{kmwa} which is somewhat underestimated, 
in our opinion. It is so large because rather high energies are required 
to excite resonances. However, it is still smaller than the radiative loss.

To conclude, we have estimated such parameters of the nuclear matter in 
heavy-ion collisions as its nuclear index of refraction, the density of 
partons, their free path length and energy loss. The RHIC data on the 
two-bump structure of the hadron distribution in away-side jets have been 
used within the notion of Cherenkov gluons. In accordance with previous 
estimates our results show that the density of partons in high energy 
heavy-ion collisions is very high. This favors the hypothesis about a 
new state of dense matter created in heavy-ion collisions.

\begin{acknowledgments}

I thank B.M. Bolotovsky for fruitful discussions.
This work has been supported in part by the RFBR grants 
04-02-16445-a, 04-02-16333.
\end{acknowledgments}

\end{document}